\documentclass[conference]{IEEEtran}
\IEEEoverridecommandlockouts
\usepackage{cite}
\usepackage{amsmath,amssymb,amsfonts}
\usepackage{algorithmic}
\usepackage{graphicx}
\usepackage{textcomp}
\usepackage{xcolor}
\def\BibTeX{{\rm B\kern-.05em{\sc i\kern-.025em b}\kern-.08em
    T\kern-.1667em\lower.7ex\hbox{E}\kern-.125emX}}

\usepackage[ruled,vlined]{algorithm2e}
\usepackage{comment}
\usepackage{dblfloatfix}
\begin{document}

\title{PoRCH: A Novel Consensus Mechanism for Blockchain-Enabled Future SCADA Systems in Smart Grids and Industry 4.0
}

\author{\IEEEauthorblockN{
Md Tamjid Hossain,
Shahriar Badsha,
Haoting Shen
    }
    
\IEEEauthorblockA{University of Nevada, Reno, NV, USA}

Email: mdtamjidh@nevada.unr.edu,\{sbadsha, hshen\}@unr.edu
}



\IEEEoverridecommandlockouts
\IEEEpubid{\makebox[\columnwidth]{978-1-7281-9615-2/20/\$31.00~\copyright2020 IEEE \hfill} \hspace{\columnsep}\makebox[\columnwidth]{ }}

\maketitle

\IEEEpubidadjcol

\begin{abstract}
Smart Grids and Industry 4.0 (I4.0) are neither a dream nor a near-future thing anymore, rather it is happening now. The integration of more and more embedded systems and IoT devices is pushing smart grids and I4.0 forward at a breakneck speed. To cope up with this, the modification of age-old SCADA (Supervisory Control and Data Acquisition) systems in terms of decentralization, near-real-time operation, security, and privacy is necessary. In this context, blockchain technology has the potential of providing not only these essential features of the data acquisition process of future SCADA systems but also many other useful add-ons. On the other side, it is evident that various type of security breach tends to take place more during any economic turmoil. These can cause even more serious devastation to the global economy and human life. Thus, it is necessary to make our industries robust, automated, and resilient with secured and immutable data acquiring systems. This paper deals with the implementation scopes of blockchain in the data acquisition part of SCADA systems in the area of the smart grid and I4.0. There are several consensus mechanisms to support blockchain integration in the field cryptocurrencies, vehicular networks, healthcare systems, e-commerce, etc. But little attention has been paid for developing efficient and easy to implement consensus mechanisms in the field of blockchain-enabled SCADA systems. From this perspective, a novel consensus mechanism, which we call PoRCH (Proof of Random Count in Hashes), with a customized mining node selection scheme has been proposed in this paper. Also, a small-scale prototype of a blockchain-enabled data acquisition system has been developed. The performance evaluation of the implemented prototype shows the benefits of blockchain technology.
\end{abstract}


\section{Introduction}
From 2011, when it first appeared in a high-tech strategy project of the German government, till today Industry 4.0 (simply I4.0 or I4.0) has crossed a long path to become a realization from only a concept. To fulfill the objective of creating self-optimized, self-cognitive, and self-customized industries, I4.0 has been incorporated with CPS (Cyber-Physical Systems), IoT (Internet of Things), IoS (Internet of Service) and Smart factory. However, one thing which is common and very important to all of these new technologies to be interconnected and updated for the successful operation of I4.0 is the continuous, secure, and automated flow of data. At the same time, the data acquisition process needs to be trustworthy and faster without the presence of a third-party to prevent unauthorized data manipulation. On the other hand, blockchain technology can be described as an open and distributed ledger that records transactions between two parties efficiently and in a verifiable and permanent way \cite{b1}. By design, blockchain is immutable and offers non-repudiation. For these certain characteristics and potentiality, nowadays blockchain technology has become very popular to be implemented in the data acquisition part of the future SCADA systems of the smart grid industry. This can be a giant leap towards the fourth industrial revolution. The motivations behind this research are –
\begin{itemize}
    \item The data acquisition process needs to be secured, automated, quick, error-free, and robust for any SCADA system which is used in smart grids and other industries.
    \item Blockchain technology can help the data acquisition process to be a smart fit in the context of I4.0
    \item Traditional consensus mechanisms of blockchain technology cannot serve efficiently for all the data acquisition process. Hence, sometimes, a customized consensus mechanism is necessary for a private and permissioned type of blockchain network
    \item Earlier researches related to the mining node selection process can be impractical in real-life scenarios. So, a random and fair mining node selection process is necessary
\end{itemize}

This major focus point of this paper is on a very important part of the blockchain technology - the consensus mechanism, from the perspective of the data acquisition process of future smart grids. The paper also discusses other essential steps of a private blockchain creation including, but not limited to, hashing mechanism, block creation, verification, and addition. Moreover, a customized mining node selection procedure has been proposed, designed, and demonstrated in the context of the SCADA system of a smart grid. The proposed PoRCH mechanism counts a random number appearance in the hash value of the measurement data and selects the node/server as a mining node that has the largest count. In case of not having a single node bearing the largest count of random number appearance, the scheme proposes to use cryptographically secure random node selection algorithm for mining purpose from all available nodes. In this way, the mechanism ensures randomness and fairness in the mining node selection process of a private blockchain network. The performance evaluation shows that the entire process requires very low computational overhead while preserving data security, privacy, and trust. In a nutshell, the major contributions of the paper are –
\begin{itemize}
    \item The paper discusses the previous research works on the consensus mechanism of the blockchain-enabled data acquisition system and points out the research gaps
    \item It bridges one of the common and popular ICS (Industrial Control System) protocols, DNP3 (Distributed
    Network Protocol 3), with the blockchain network
    \item The paper proposes a novel consensus mechanism called PoRCH and present a simplified demo by including a customized mining node selection procedure for a private and permissioned blockchain-enabled data acquisition system where incentive or penalty is not required for the validators/miners. 
\end{itemize}

In the remainder of this paper, we have used the words – 'nodes'', 'servers', 'field devices', 'relay servers' alternately depending upon the context, but they are implemented as one type of entity that can both transmit and receive data.

\section{Related works}
In \cite{b2}, blockchain has been applied to the ICS network of device nodes where every device node executes the smart contract and records the transaction data. Any device nodes on the ICS network can record all transaction data. However, it may then possible for an adversary to compromise a device node and manipulate the data. So, there has to be a novel and decentralized way of tackling cyber-attacks before they take place and cause a single-point-of-failure and data manipulation. Taking this into account, researchers of  \cite{b23} propose to enhance the integrity and confidentiality of synchrophasor communication networks in an ICS environment through blockchain and essential cryptographic tools (e.g. bloom filter, ECDSA, etc.). The researchers of \cite{b3} proposes a scheme- BUS where IoT devices send encrypted data to UAV (Unmanned Aerial Vehicle), and UAV validates the sender's identity using a bloom filter. UAV then sends the data to the nearest server who prepares the data to add to the blockchain. Several other validators also validate the block before adding it to the blockchain. In other researches, blockchain has been applied towards data security in ICS \cite{b4,b5}. Reference \cite{b4} uses ethereum blockchain in its prototype implementation and may suffer from the problems of ethereum blockchain such as scalability and latency.
 
Reference \cite{b5} uses a customized mining node selection procedure that follows the process of selecting a node as a mining node if its energy consumption data is the closest to the average. However, in a practical scenario, the electricity consumption behavior of any user may not change rapidly and remain the same at least for a while. Thus, it may possible for any node to have the almost same data for concurrent time slots and so, there may be a node whose data will be closest to average for these concurrent time slots. As a consequence, a node may remain as a mining node for a considerable amount of time.
If an adversary, somehow, finds out the mining node and compromise it, he/she can exploit this time slots to design and execute his/her attack. Reference \cite{b6} addresses the problem of reaching consensus in distributed social networks. However, they have not incorporated the mining technique, rather they've used wired connected RSUs (roadside units). The consensus mechanism is lightweight but may not be applied to the blockchain-based smart grid data acquisition system or any system that requires a near-real-time operation through wireless connectivity. 
A novel consensus mechanism named PoAh (Proof-of-Authentication) has been proposed in \cite{b8}. This consensus mechanism introduces the trusted party in the validation of blocks and solves many problems of traditional consensus mechanisms. However, these schemes have not bridged the OT protocols with the communication network of the blockchain which in turn, has motivated the authors to proposed the PoRCH scheme that aims to connect the dots in this domain.

\section{Main Design}
\subsection{Communication Protocol}
The communication system of an ICS can be broadly categorized into IT (Information Technology) and OT (Operational Technology).
IT systems use ERP systems that mostly uses SMTP, SNMP, OPC, SMB, HTTP, XML protocols whereas OT systems encompass MES, DCS, SCADA, RTUs, PLCs, sensors, etc. and use MODBUS, PROFIBUS, PROFINET, DNP3, EtherNet/IP, RS-232, etc.

\begin{figure}[ht!]
    \centerline{\includegraphics[width=\linewidth]{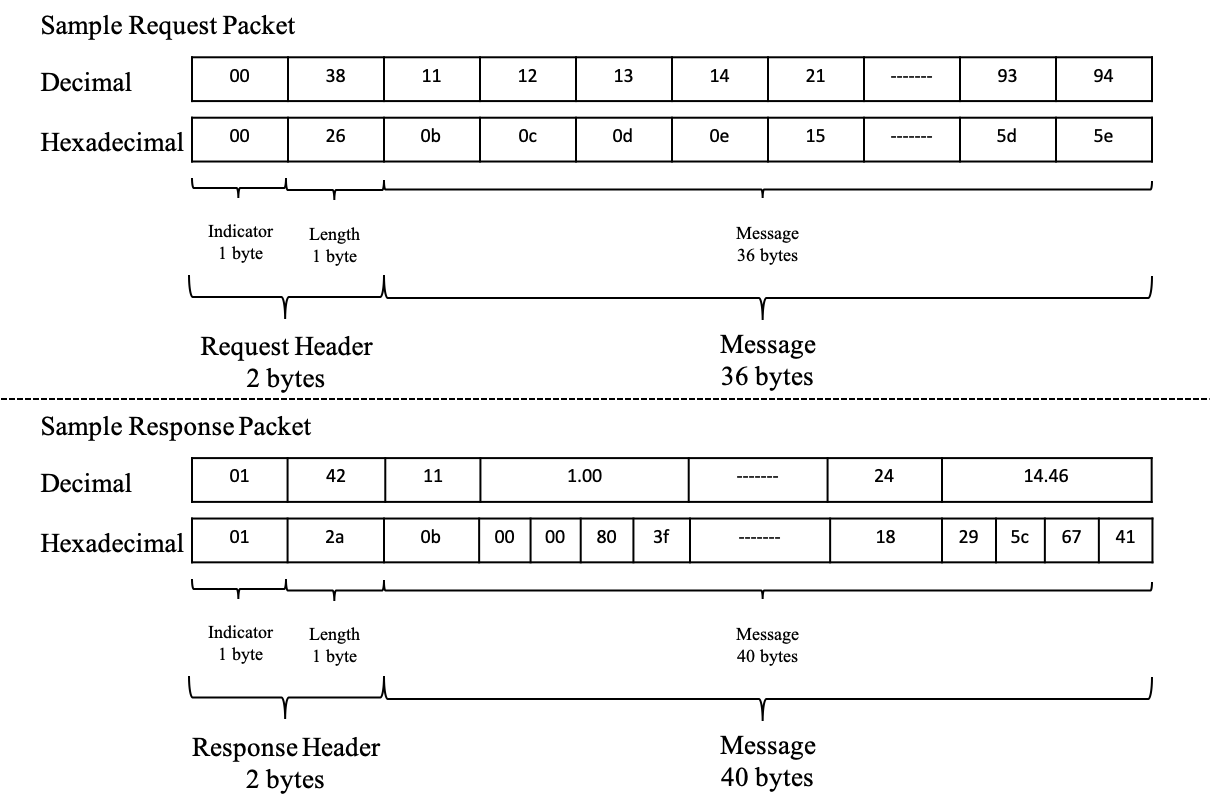}}
    \caption{Sample request and response packet.}
    \label{req_resp}
\end{figure}

Here, in this paper, DNP3m (DNP3 minus, a simplified version of DNP3) protocol has been used to collect the data by the control center and data aggregator node from relay nodes. It is widely used in the US power grid infrastructure. It is an application layer protocol built on top of the TCP protocol \cite{b9}.
The Sample request and response packets are also depicted in Fig.~\ref{req_resp}. 
However, the goal of this paper is not to find out the most secure communication protocol and discuss in-depth but to implement blockchain technology to make the data acquisition system decentralized, secured, private, and trustworthy. In the application layer, both the request and response header contains 2 bytes. The first byte represents the identity of the protocol, i.e. the protocol is DNP3m. In this implementation, the first byte is set to ‘0’ for request header and ‘1’ for response header. The second byte reflects the length of the entire message. Then, the message is incorporated with the headers.

\subsection{Relay Servers}
Relay servers are the primary sources of measurement data. They are associated with different field devices, sensors, actuators, PMUs, valves. In the prototype demonstration, four nodes/servers have been used to represent the relay servers. The relay servers can act both as servers and clients based on operational needs. Here, relay servers use the DNP3m protocol to send the response with measurement data. UTF-8 (8-bit Unicode Transformation Format) has been used as an encoding mechanism for sending string responses. Another important task of the relay servers is to take part in the voting process of mining node selection. To cast vote for the appropriate node during the mining node selection process, the relay servers first use hashing technique to hash their measurement data. Then, they count the appearance of the random number in their hash values and finally votes for the node to be selected as the mining node who has the largest count.

\subsection{Data Aggregator (DA)}
Data Aggregator (DA) plays the most vital role in the proposed data acquisition system. It is responsible for collecting the data, then preprocessing it and coordinate mining node selection and all kinds of the verification process. DA should preferably reside inside the internal network with the relay servers for efficient operation and low computational overhead and time. In this demonstration, DA works as both in server and client mode. As soon as it gets a data acquisition request from CC (Control Center), it starts its operation to collect the data from relay servers and finally completes its operation for a cycle by sending updated blockchain to CC and other relay servers. Another important task of DA is to generate random numbers for assisting the mining node selection procedure. Moreover, it might be the case to have multiple nodes having the largest and same appearance counts of the random number in their hash values or the case to have all nodes having the count as zero (0). In these scenarios, the DA delegates one node from all available nodes based on a random selection algorithm.

\subsection{Control Center (CC)}
The CC is responsible for initializing and controlling the operation. CC can be any physical or cloud server and may have HMI (Human Machine Interface) attach with it for better user experiences. At first, CC sends the data acquisition command to DA and finally receives the updated copy of the blockchain containing measured data at the end of each cycle. For simplicity, here in this demonstration, each data acquisition cycle has been considered as 15 seconds. Of course, this period is configurable according to the requirement of the data acquisition system, data generation, and processing time.

\subsection{Blockchain}
Blockchain is an open, distributed, and immutable ledger system that was first conceptualized as a public transaction ledger of the cryptocurrency bitcoin by Satoshi Nakamoto in 2008. Since then, blockchain technology has been proposed to apply in several domains, including but not limited to, task allocation \cite{b19}, ride-sharing \cite{b20}, cybersecurity information sharing \cite{b21}, cyber insurance \cite{b22} etc. In this paper, we have considered extending the application of blockchain in the smart grid area.

\begin{figure}[ht!]
    \centerline{\includegraphics[width=\columnwidth]{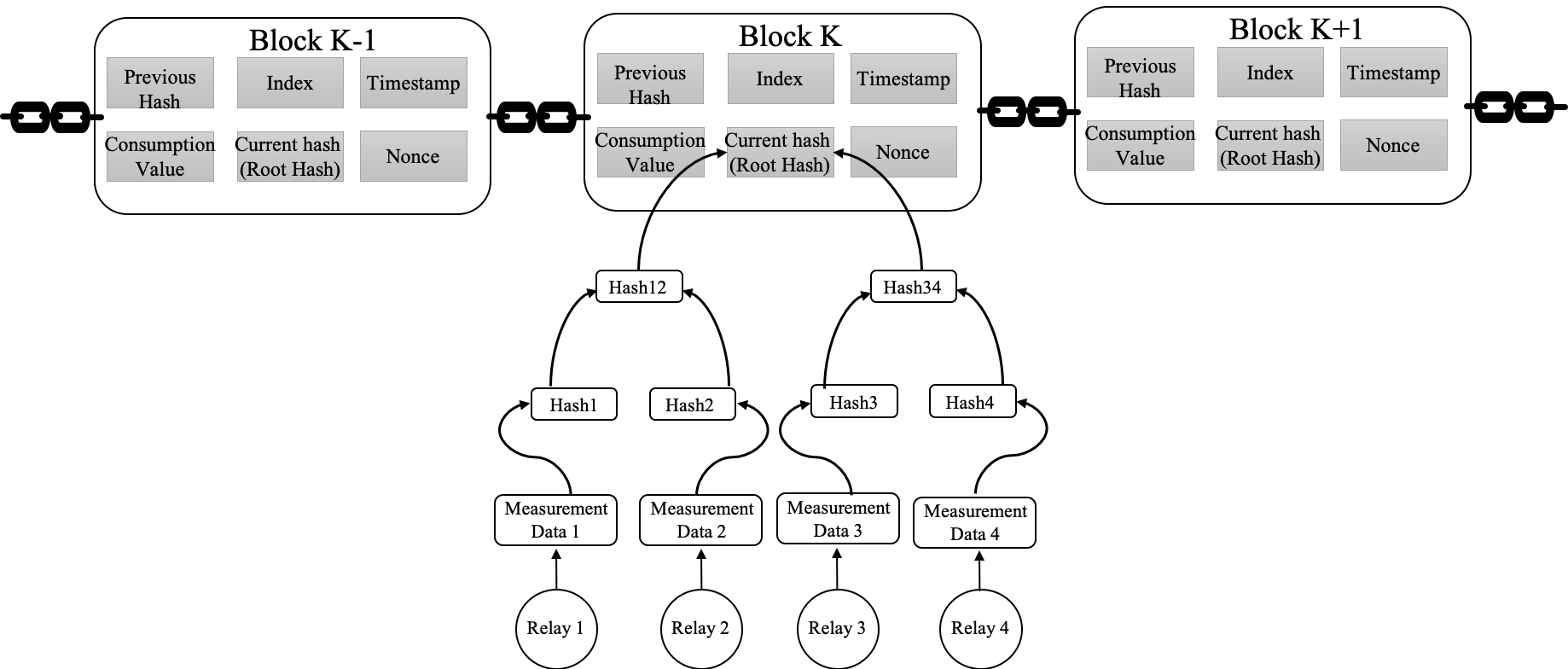}}
    \caption{Example blockchain.}
    \label{exam_blockchain}
\end{figure}

Blockchain implementation starts with the selection of a mining node and ends with the addition of new blocks in a repetitive manner. In this part, data preprocessing, data aggregation, existing blockchain verification, hashing, new block creation (see Fig.~\ref{exam_blockchain}), block verification, block addition occurs as an intermediate process. 
Design of the proposed blockchain can be well described by the following processes.\\

\subsubsection{Hashing Mechanism}\label{HM} It refers to a process of generating an output of a fixed or predefined length regardless of the length of the input. The result of this computational function is called a hash. In a blockchain, the process of using a specific hash function to process a transaction or data is called hashing.

Rarely, a hash collision may occur as a result of the ‘Birthday Box’ incident \cite{b13}. A hash collision is defined as an instance in which two or more observations are hashed to the same value \cite{b14}. In this demonstration, SHA-256 has been implemented as a hashing algorithm by using Python 3.0 ‘hashlib’ module.\\
\begin{figure*}[b]
    \centerline{\includegraphics[width=\textwidth]{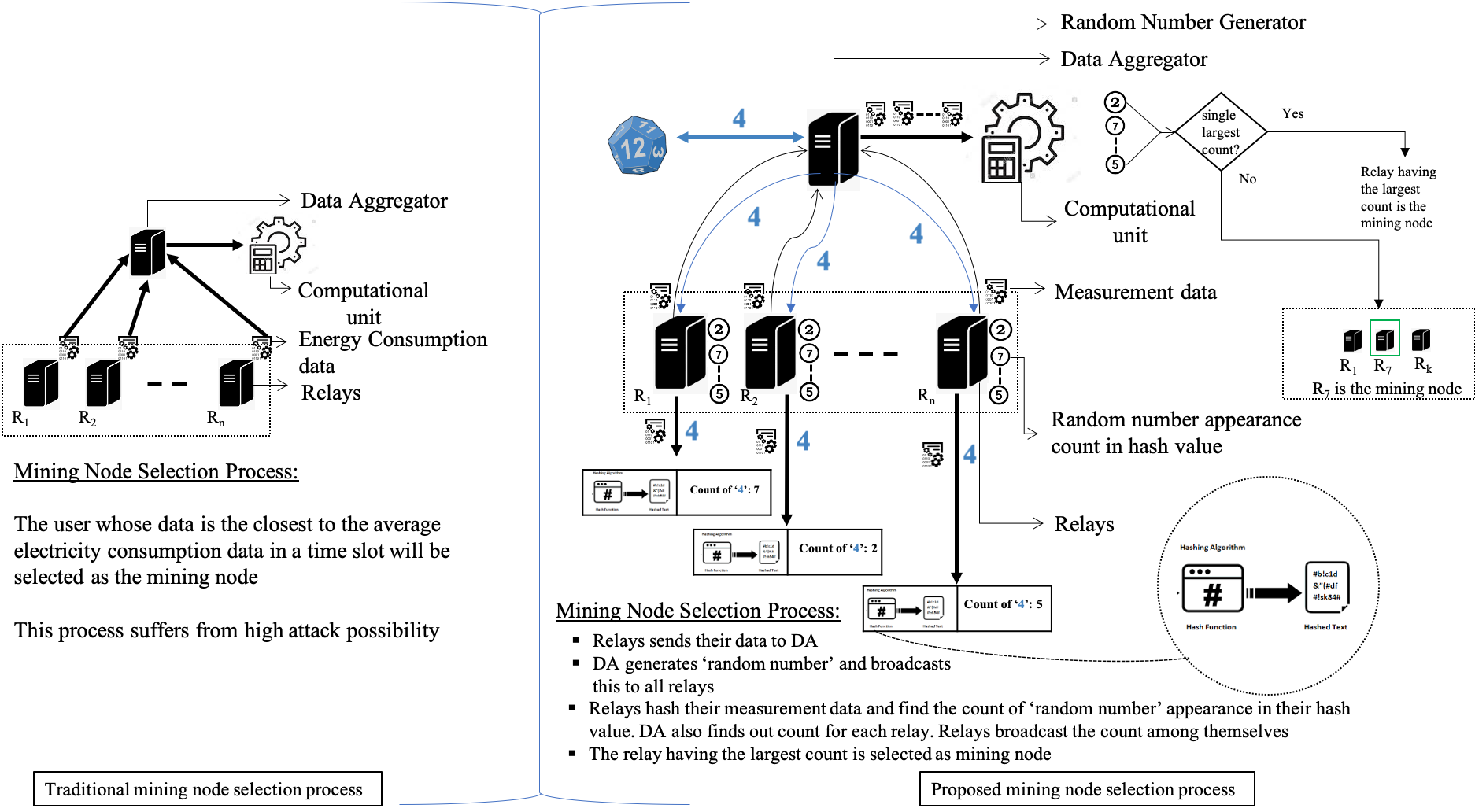}}
    \caption{Traditional vs proposed mining node selection procedure.}
    \label{mining_node_select}
\end{figure*}
\subsubsection{Consensus Mechanism} At a high level, a consensus mechanism can be described as the process of adding newly published blocks in the blockchain after testing the validity by validators/miners to ensure the trust in the network. Over the years, numerous consensus mechanisms have been proposed, developed, and implemented in several blockchain networks. Both public and private blockchain use consensus mechanism. Popular consensus mechanisms are PoW (Proof of Work), PoS (Proof of Stake), DPoS (Delegated Proof of Stake), PBFT (Practical Byzantine Fault Tolerance), PoA (Proof of Authority), RAFT \cite{b15}, etc. Each one has its advantages and drawbacks. PoW requires high computational overheads, PoS is biased to the wealthiest validators and thus unfair for new participants, DPoS is less decentralized and less resilient, PBFT only works on a permissioned blockchain due to the lack of anonymity \cite{b16}. Though the authors of \cite{b24} have proposed a consensus mechanism named PoD (Proof of Driving) which aims to improve PBFT in VANET (Vehicular Ad hoc Network) application, they have mainly focused on implying it in a public blockchain network. However, in a private blockchain that may be adopted by a small industry and where any incentive or penalty mechanism may not necessary, above mentioned consensus mechanisms would not suit properly due to their drawbacks. So, some sort of customized and lightweight consensus mechanism like the proposed PoRCH mechanism would be a better choice.
\begin{algorithm}
\SetAlgoLined
    \SetKwInOut{Input}{Input}
    \SetKwInOut{Output}{Output}
    \SetKwProg{Fn}{Function}{ :}{end}
    
    \Input{Measurement Data as $D_m$, Relay Name as $N_r$, Random Number from DA  as $R$}
    \Output{The name of the mining node}
    \vspace{5pt}
    \Fn{Generate\_Hash($D_m$)}{
    $hash$ $\leftarrow$ hashlib.sha256(block).hexdigest()\\
    \Return $hash$\;
    }
    \vspace{5pt}
    \Fn{Count($hash$, $R$)}{
    $valCount$ $\leftarrow$ $hash$.count($R$))\\
    \Return $valCount$\;
    }
    \vspace{5pt}
    
    \textbf{Set Dictionary: } $R_v$ $\leftarrow$ \{$N_r: valCount$\}\\
    \textbf{Sort Dictionary: } $S_{R_v}$ $\leftarrow$ sorted($R_v$.items(), key=lambda item: item[1], reverse=True)\\
    \textbf{number of largest count: } $LarCount$ $\leftarrow$ list($S_{R_v}$.values()).count($S_{R_v}$[list($S_{R_v}$.keys())[0]])\\
    \vspace{5pt}
    \uIf{$LarCount ==$ 0}{
        DA selects MiningNode Randomly \;
    }
    \uElseIf{$LarCount ==$ 1}{
        MiningNode  $\leftarrow$ $list(S_{R_v})[0]$ \;
    }
    \uElseIf{$LarCount > $ 1}{
        DA selects MiningNode Randomly \;
    }
    \Else{
        Stop the selection process \;
    }
    \caption{Mining node selection}
\end{algorithm}
\begin{figure*}[b]
    \centerline{\includegraphics[width=\textwidth]{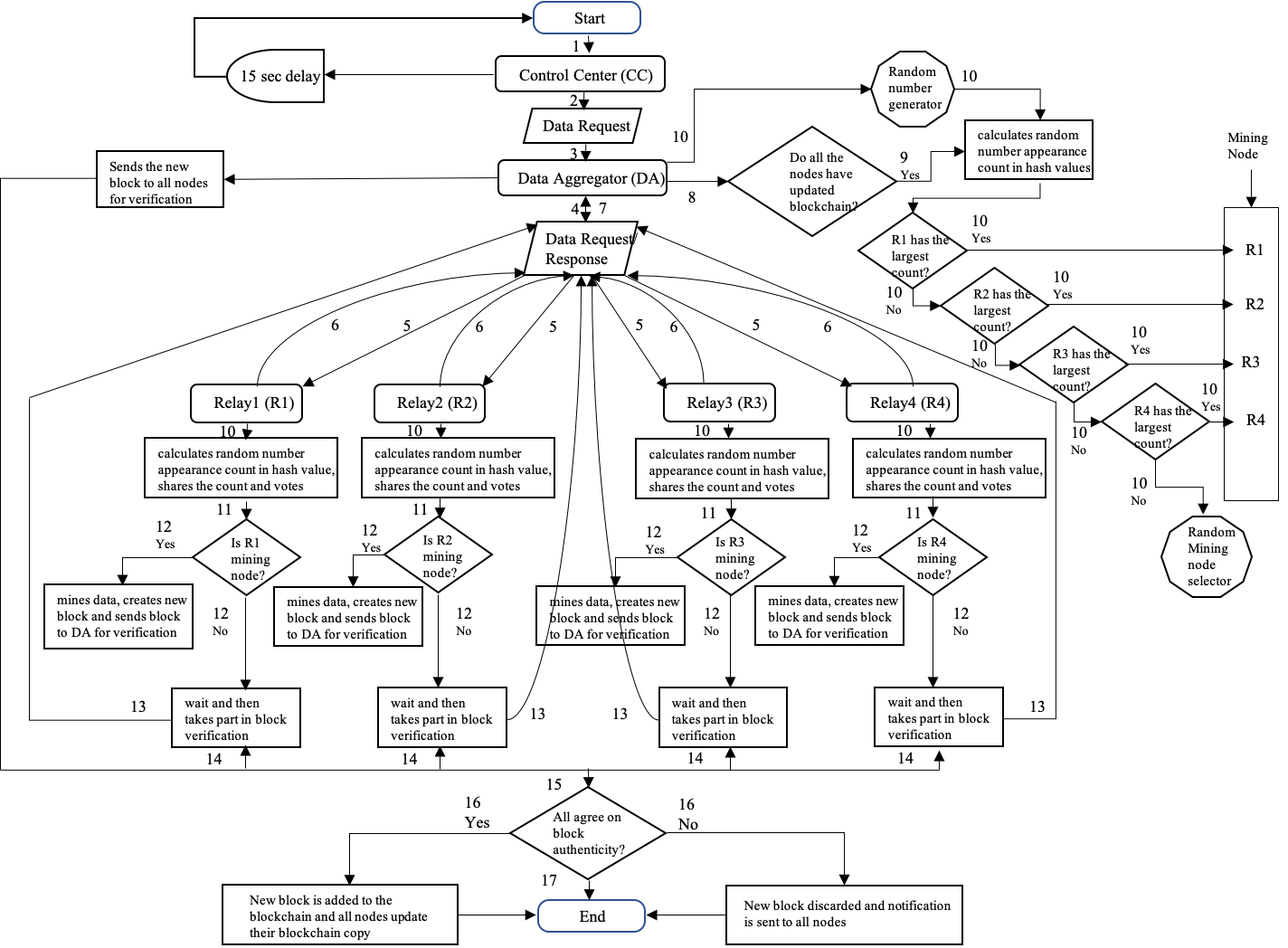}}
    \caption{Flowchart of the operational procedure of proposed scheme. The process follows the number sequence of the flowchart.}
    \label{flow}
\end{figure*}

\textit{\textbf{Mining node selection:}} Considering the drawbacks of traditional consensus mechanisms, a customized mining node selection procedure has been designed and demonstrated (see Fig.~\ref{mining_node_select} and algorithm 1). First, the DA server requests data from the relay servers. The relay servers then send the encrypted measurement data to DA. After getting the measurement data from each relay, DA generates a random number and broadcasts this random number to all the relay servers. Relay servers then hash their measurement data and find out that random number appearance count from their respective hash values. Next, the relay servers share the count among each other. Meanwhile, the DA also hashes each measurement and finds out the same counts against each hash value. In this stage, every server including the DA has the same counts. At the final stage, all the servers cast their votes for the relay server having the largest count of random number appearance. If all the relay servers reach a consensus, the DA server selects the relay having the largest count as the mining node for that cycle. However, if multiple relay servers have the same largest counts or all of them have the count as zero (0), then the DA selects the mining node randomly following a cryptographically secure random selection algorithm.  The functions $random.randint(start, end)$ and $random.choice(list)$ of the module $random$ from the python standard library has been used to generate random numbers and select random node in our simplified demonstration. However, for achieving more security and randomness, the proposed scheme can be extended to adopt CSPRNG (cryptographically secure pseudorandom number generator) algorithm. We have considered to include it in our future works.\\
\begin{figure*}[b]
    \centerline{\includegraphics[width=\textwidth]{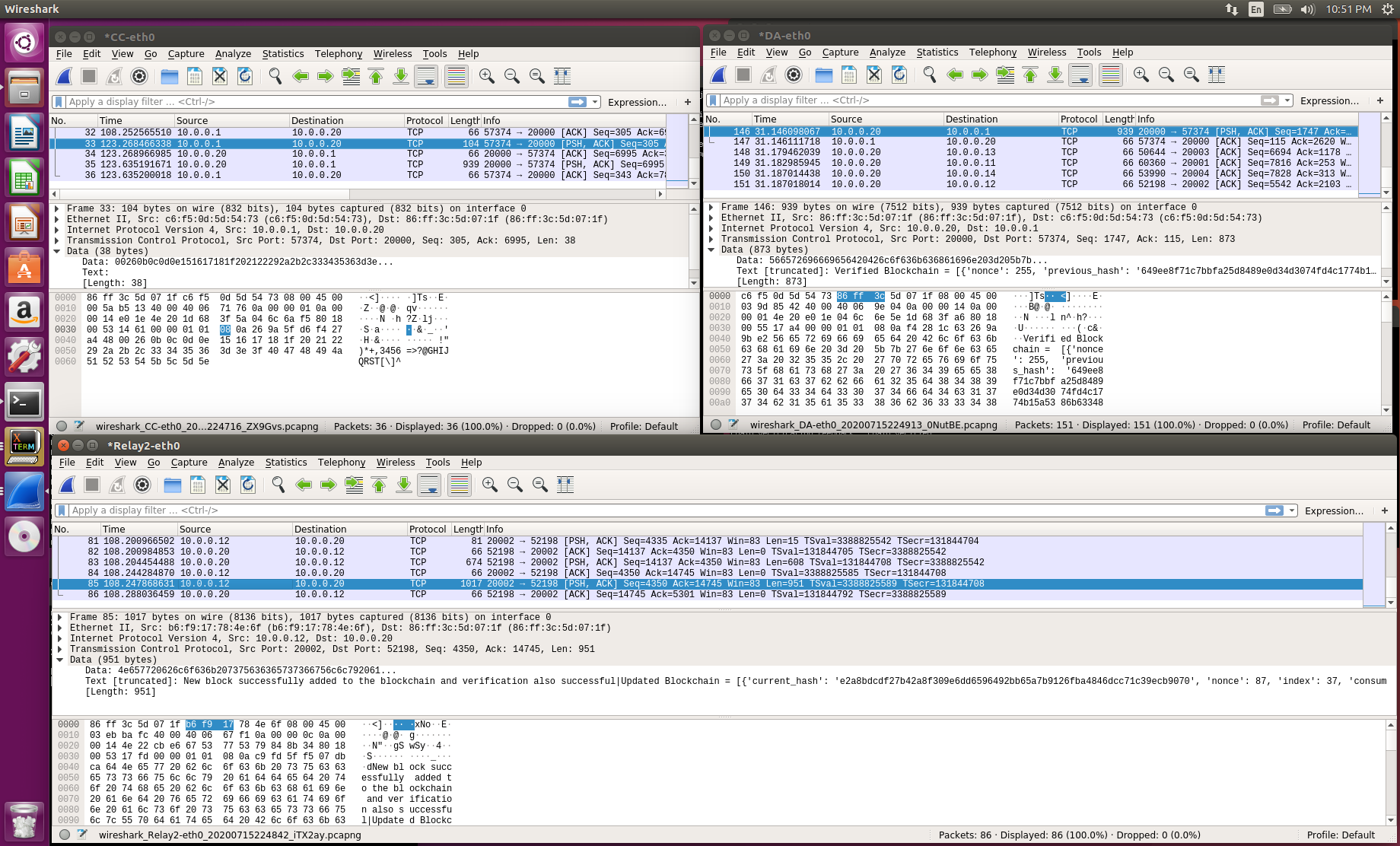}}
    \caption{Network packet analysis using Wireshark.}
    \label{wireshark}
\end{figure*}

\textit{\textbf{Block creation, verification and addition:}} The working principle of the simplified demonstration of the scheme has been described in Fig.~\ref{flow}. Here, as soon as the mining node selection is completed, the DA sends the encrypted and aggregated measurement data to the mining node. The measurement data is then hashed by the mining node using the technique described under \ref{HM}. During hashing operation, a Merkle tree structure has been followed. Then, the mining node creates a new block with the previous hash, index, timestamp, consumption value, current hash (root hash), and nonce. A typical block has been shown in Fig.~\ref{exam_blockchain}. All the relay servers and the DA participate in the block verification process. During the verification process, the servers exchange responses several times. If all parties agree on the validity of the block, the DA sends the block addition request to all the servers. Finally, after the successful addition of the block to the blockchain, the DA sends the updated blockchain to the control center. \\

\section{Performance Evaluation}
\subsection{Experimental Setup}
For experimental purposes, we set up a testbed in a virtual machine. Ubuntu 20.04 (64-bit) with a base memory of 8 GB and storage of 20 GB has been considered as the operating system. To build up servers/nodes, Mininet version 2.2.2 has been used. We have used Python 3.0 as the programming language. Moreover, We have used Wireshark (previously known as Ethereal) to capture the network packet (see Fig.~\ref{wireshark}). DNP3 runs over TCP and UDP. The Wireshark DNP dissector registers for TCP and UDP ports 20000 by default and checks whether the TCP segment it's handed has at least 2 bytes of data or not.
The measurement data is coming from an IEEE 9-bus system \cite{b17}.

\subsection{Result and Discussion}
As this is a no reward-no penalty based private and permissioned blockchain model for smart industries, miners/validators do not need to compete for mining opportunities. So, a complex mathematical problem-solving approach has been avoided which in turn saves energy that may incur due to huge computational overhead as it occurs in PoW. At the same time, the mining process needs to be random and fair which can also provide security to the data acquisition system. The proposed approach is a good fit in this context. It has the below advantages over them-
\begin{itemize}
    \item One of the goals of the paper is to find out an efficient consensus mechanism for private blockchain which is suitable to implement in SCADA systems in the context of smart grids. So, incentive or penalty mechanisms have been avoided in the proposed scheme. This, in turn, solves the huge computational overhead or unfair selection issues of PoW or PoS mechanisms
    \item A level of fairness has been ensured as all the relay servers get the chance of voting. At the same time, the random number generator and the random node selector ensure the randomness of the mining node selection process
    \item As all the servers have the same counts, even if somehow, an adversary compromises a server, he/she cannot alter the count to become a mining node without revealing his/her identity. This provides security in mining node selection process
\end{itemize}

However, questions may arise regarding the complexity and computational time for this process. It is desired to be less time consuming, especially for the operations where near-real-time data is very important. Surely, selecting the mining node through a simple random process by the DA would reduce both the complexity and computational time. But, in that case, an adversary may easily become a mining node by only compromising the DA. So, a trade-off has been made between the computational time, complexity, and security.

\begin{figure}[ht!]
    \centerline{\includegraphics[width=\linewidth, height = 4.7cm, keepaspectratio]{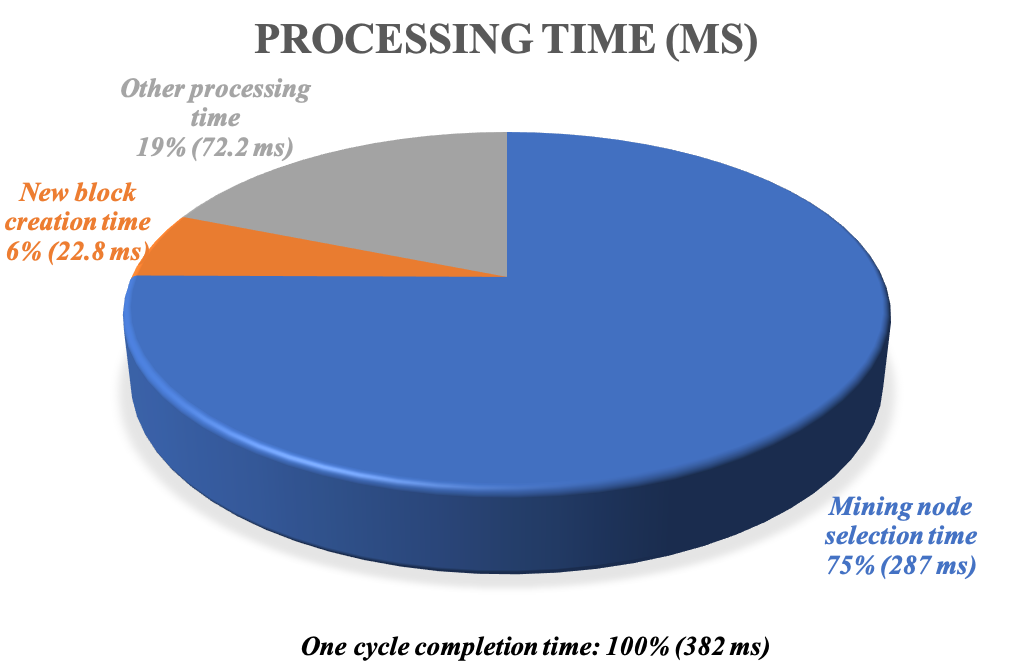}}
    \caption{Processing time of the prototype.}
    \label{eval}
\end{figure}

The simplified experiment shows that an entire operation cycle with four relay servers and four measurement data (one from each server) per block takes only 382 milliseconds. Mining node selection procedure takes around 75\% of the processing time of a cycle (see Fig.~\ref{eval}).\\

If the number of nodes increases, the processing time would also increase. One possible way to overcome this is to allow only a subset of total nodes, say for an example, $K$ nodes to be eligible as a mining node among $N$ nodes where $K < N$ and $K \subset N$. Then, at a certain point, the increment of mining node selection time would be halted even if the number of nodes increment goes on. Other possible measures can be taken to make the data acquisition process more secure and trustworthy, e.g., data encryption, data anonymization using pseudonyms, node identity anonymization. Moreover, faster authentication could be achieved by using a bloom filter. There is a very low possibility of hash collision during the hashing mechanism.
Reference \cite{b5} proposed to decrease the probability of hash collision by increasing the array size of the bloom filter. There is a debate on using double-SHA-256 to overcome the hash collision but bitcoin is using it as $SHA256(SHA256(x))$ \cite{b18}, where $x$ is the block\_header. 

\section{Conclusion}
In this paper, a novel consensus mechanism named PoRCH with a personalized mining node selection process has been developed from a practical point of view for the SCADA systems in the field of smart grids and Industry 4.0. The scheme bridges the OT protocol, DNP3 to the blockchain technology while maintaining randomness and fairness of the process. The paper also presents a simplified and lightweight demonstration of the proposed scheme. The demonstration verifies the low computational overhead requirement of the scheme and reflects the characteristics of a blockchain model.

\end{document}